\documentstyle[11pt]{article}

\newlength{\bredde}
\def\slash#1{\settowidth{\bredde}{$#1$}\ifmmode\,\raisebox{.15ex}{/}
\hspace*{-\bredde} #1\else$\,\raisebox{.15ex}{/}\hspace*{-\bredde} #1$\fi}
\newcommand{\beq}{\begin{equation}}
\newcommand{\eeq}{\end{equation}}
\newcommand{\ba}{\begin{array}{ccc}}
\newcommand{\ea}{\end{array}}
\newcommand{\nn}{\nonumber}
\newcommand{\noi}{\vspace{12pt}\noindent}
\newcommand{\lG}{\raise.3ex\hbox{$\stackrel{\leftarrow}{G}$}}
\newcommand{\lU}{\raise.3ex\hbox{$\stackrel{\leftarrow}{U}$}}
\newcommand{\lP}{\raise.3ex\hbox{$\stackrel{\leftarrow}{{\cal P}}$}}
\newcommand{\leta}{\raise.3ex\hbox{$\stackrel{\leftarrow}{\eta}$}}
\newcommand{\lOmega}{\raise.3ex\hbox{$\stackrel{\leftarrow}{\Omega}$}}
\newcommand{\ldr}{\raise.3ex\hbox{$\stackrel{\leftarrow}{\delta^r}$}}
\def\m2{{\mathcal{M}}^{\dagger}{\mathcal{M}}}
\def\mb2{M^2}

\def\beqn{\begin{eqnarray}}
\def\eeqn{\end{eqnarray}}

\def\gtwid{\raise.3ex\hbox{$>$\kern-.75em\lower1ex\hbox{$\sim$}}}
\def\ltwid{\raise.3ex\hbox{$<$\kern-.75em\lower1ex\hbox{$\sim$}}}

\def\la{\lambda}

\def\Tr{ {\rm Tr} }

\begin{document}
\topmargin -1.4cm
\oddsidemargin -0.8cm
\evensidemargin -0.8cm
\title{\Large{{\bf Spectral Sum Rules of the Dirac Operator and \\ Partially 
Quenched Chiral Condensates}}}

\vspace{1.5cm}

\author{~\\{\sc P.H. Damgaard} and {\sc K. Splittorff}\\~\\
The Niels Bohr Institute\\ Blegdamsvej 17\\ DK-2100 Copenhagen {\O}\\
Denmark}
\date{\today} 
\maketitle
\vfill
\begin{abstract} 
Exploiting Virasoro constraints on the effective finite-volume partition 
function, we derive generalized Leutwyler-Smilga spectral sum rules of the 
Dirac operator to high order. By introducing $N_v$ fermion species of
equal masses, we next use the Virasoro constraints to compute two (low-mass 
and large-mass) expansions of the partially quenched chiral condensate 
through the replica method of letting $N_v \to 0$. The low-mass expansion can
only be pushed to a certain finite order due to de Wit-'t Hooft poles, but
the large-mass expansion can be carried through to arbitrarily high order.
Results agree exactly with earlier results obtained through both Random
Matrix Theory and the supersymmetric method. 
\end{abstract}
\vfill

\begin{flushleft}
NBI-HE-99-52 \\
hep-th/9912146
\end{flushleft}
\thispagestyle{empty}
\newpage

\section{Introduction}

\noi
The exact spectral sum rules for the Dirac operator in finite-volume gauge 
theories derived by Leutwyler and Smilga \cite{LS} and later by Smilga
and Verbaarschot \cite{SmV} have sparked a huge
activity in the field. It has turned out that there is an exact relation
between universality classes in Random Matrix Theories and the three
main ways in which chiral symmetry can break spontaneously, depending on
the gauge group and the representation of the fermion fields 
\cite{SV,V,ADMN}. The Random Matrix Theory formulation brings
a completely different set of techniques into play, and much has been
learned from this approach. One of the main efforts recently has nevertheless
been to see how all the results obtained from Random Matrix Theory can
be derived directly from the effective partition function. One of
the first steps in this direction was a set of very compact relations that
expressed microscopic spectral correlators (including the microscopic
spectral density itself) directly in terms of effective partition functions
with additional quark species \cite{AD}. More recently, many of these 
results have been derived directly from the effective field theory \cite{OTV}.
The idea has been to 
compute the (partially) quenched chiral condensate and higher chiral 
susceptibilities through an appropriately
extended effective Lagrangian, and from this derive the microscopic
spectral correlators. As the technique of ref. \cite{OTV} makes use of
additional quark species, of which half are bosonic and the others are 
fermionic, this extends the flavor symmetry to a super
Lie group at intermediate steps. For this reason it is commonly known as
the supersymmetric method (although it, as applied, has nothing to do with
space-time supersymmetry).

\noi
The purpose of this paper is to explore another technique that can be used
to derive the partially quenched chiral condensate. This is based on the
so-called replica method\footnote{Or replica ``trick'', a name that 
suggests magic and trickery. We prefer the more neutral terminology.} 
in which one extends the theory with $N_v$ 
additional fermionic species, and takes the limit $N_v \to 0$ at the end
of the calculation. The replica method is known in other contexts to be
often problematical \cite{VZ} (see however also the recent discussion
on this issue \cite{Mezard}), but we shall find no fundamental difficulties
with this technique in the present context.  

\noi
Let us first outline how the replica method can be used to compute partially
quenched averages in QCD. Consider the QCD partition function with $N_f$
physical quark fields and $N_v$ additional quark fields of degenerate
masses $m_v$. We restrict ourselves to gauge field sectors of fixed 
topological charge $\nu$, which we for simplicity from now on take to be
non-negative:
\beq
{\cal Z}_{\nu}^{(N_{f}+N_{v})} ~=~ 
\left(\prod_{f=1}^{N_{f}} m_f^{\nu}\right)m_{v}^{N_{v}\nu}
\int\! [dA]_{\nu}
~{\det}'(i\slash{D} - m_v)^{N_{v}}\prod_{f=1}^{N_{f}}
{\det}'(i\slash{D} - m_f) ~e^{-S_{YM}[A]} ~, \label{Zoriginal}
\eeq
where the determinants are taken over non-zero modes only.
This can be considered as an average over gauge fields (and physical
fermions, here already integrated out) of $N_v$ identical replicas of the 
fermionic partition function
\beq
{\cal Z}_v ~\equiv~ \int\! d\bar{\psi}d\psi~ \exp\left[\int\! d^4 x 
\bar{\psi}(i\slash{D} - m_v)\psi\right] ~,
\eeq
in the sense that
\beq
{\cal Z}_{\nu}^{(N_{f}+N_{v})} ~=~ \left(\prod_{f=1}^{N_{f}} m_f^{\nu}\right)
\int\! [dA]_{\nu}\prod_{f=1}^{N_{f}}{\det}'(i\slash{D} - m_f)
\left[{\cal Z}_v\right]^{N_v} ~e^{-S_{YM}} ~.
\eeq
In condensed matter physics this (unnormalized) average is conventionally
denoted by a bar:
\beq
{\cal Z}_{\nu}^{(N_{f}+N_{v})} ~=~ \overline{[{\cal Z}_v]^{N_{v}}} ~.
\eeq
The subscript $v$ will throughout denote ``valence'', and the additional 
quarks are thus valence quarks, a terminology borrowed from lattice gauge 
theory. Obviously, if we let $N_v=0$ we simply recover the original QCD 
partition function. For $N_f + N_v$ not too large, the theory is presumed
to undergo, in the chiral limit of massless quarks, spontaneous chiral 
symmetry breaking according to the usual
pattern SU($N_f+N_v)\times$SU($N_f+N_v) \to$ SU($N_f+N_v$). Taking the
limit $N_v\to 0$ must therefore incorporate some kind of analytic
continuation in $n$ of the Lie group SU($n$). 

\noi
For any fixed $N_v$ we can compute observables
in the extended theory by viewing the partition function as the generating
function of the $n$-point functions of $\bar{\psi}\psi$. We shall here focus
on just the chiral condensate itself, and only on the chiral condensate of
the additional fermionic copies. As we will take $N_v\to0$, this becomes the
partially quenched condensate in the theory with $N_f$ physical fermions.
We define this partially quenched (mass-dependent) chiral condensate by
\beq
\frac{\Sigma_{\nu}(\mu_v,\{\mu\})}{\Sigma} ~\equiv~ \lim_{N_{v}\to 0}
\frac{1}{N_{v}}\frac{\partial}{\partial \mu_{v}}
\ln {\cal Z}_{\nu}^{(N_{f}+N_{v})} ~, \label{firstdef}
\eeq
where $\Sigma$ is the physical infinite-volume chiral condensate in the
theory with $N_f$ fermions, $\mu_j\equiv m_jV\Sigma$ and similarly
$\mu_v\equiv m_vV\Sigma$. Treating $N_v$ as a
parameter that is not
restricted to be integer (which is permitted by the representation
of the partition function in eq. (\ref{Zoriginal})), we can make a Taylor
expansion in $N_v$:
\beq
{\cal Z}_{\nu}^{(N_{f}+N_{v})} ~=~ {\cal Z}_{\nu}^{(N_{f})} + 
\left.N_{v}\frac{\partial}{\partial N_{v}}
{\cal Z}_{\nu}^{(N_{f}+N_{v})}\right|_{N_{v}=0} + \ldots
\eeq
and thus write, equivalently,\footnote{One can explicitly check that the
two derivatives commute in our case.}
\beq
\frac{\Sigma_{\nu}(\mu_v,\{\mu\})}{\Sigma} ~=~ 
\left.\left[{\cal Z}_{\nu}^{(N_{f})}\right]^{-1}\frac{\partial}{\partial
N_{v}}\frac{\partial}{\partial \mu_{v}} 
{\cal Z}_{\nu}^{(N_{f}+N_{v})}\right|_{N_{v}=0} ~.
\eeq
The partially quenched chiral condensate is a particularly convenient
quantity for lattice gauge theory simulations, and comparisons with theory
have already been made both in the topologically trivial sector \cite{VMC}
and, very recently, in sectors of non-vanishing topological charge
$\nu$ \cite{DEHN}. We note here that higher-order partially quenched
chiral susceptibilities, which are also readily studied by Monte Carlo
simulations, can be derived in a completely analogous manner. For a chiral
$k$-point function one simply needs to introduce $N_{v1},\ldots,N_{vk}$
sets of different additional species (each of $N_{vi}$'th degenerate
masses $\mu_{vi}$), and take the combined limit of all $N_{vi}\to 0$ in
the end, after having performed the required differentiations. 

\noi
In this paper we will show how the replica method can be applied to
the effective finite-volume partition function, 
and in this way we shall derive
analytical series expansions for the partially quenched chiral
condensate. As a by-product of the analysis we will get, for free,
generalized Leutwyler-Smilga sum rules to very high order.
The plan of the paper is as follows. In the next section we review the
technique for deriving high-order expansions of the effective partition
function on the basis of an iterative sequence of partition function
constraints. From this expansion we immediately derive a long list of
generalized spectral sum rules of the Dirac operator. In section 3 we
turn to the replica method, where we first exploit the same partition
function constraints to derive a small-mass expansion for the partially
quenched chiral condensate. We also show how a different set of
partition function constraints can be used to derive a very high order
expansion in large masses.  In section 4 we point out that the usual
Leutwyler-Smilga spectral sum rules, which by definition are taken
with respect to the massless theory, have a natural generalization where
the spectral sums are taken with respect to the theory with massive
fermions. We derive a series of such massive spectral sum rules, 
perturbatively expanded in the physical masses. In section 5 we discuss
some generalizations to different patterns of spontaneous chiral symmetry
breaking, and point out why the present series expansions
cannot be used to derive the microscopic spectral density of the Dirac
operator itself. In section 6 we briefly list the main conclusions.

\setcounter{equation}{0}
\section{Spectral Sum Rules from Virasoro Constraints}
\label{masslessSR}

\noi
Spectral sum rules of the Dirac operator are derived by comparing the
full QCD partition function with the effective low-energy partition function.
In the large-volume scaling region $V \ll 1/m_{\pi}^4$, this effective
partition function is to leading order \cite{LS}
\beq
{\cal Z}^{(N_f)}_{\nu} ~=~ \int_{U\in U(N_f)}\! dU~ (\det U)^{\nu}
\exp\left[\frac{1}{2} {\rm Tr}({\cal M}
U^{\dagger} + U{\cal M}^{\dagger})\right] ~, \label{ZUE}
\eeq
where ${\cal M}$ is the quark mass matrix rescaled by the space-time 
volume $V$ times the
value $\Sigma$ of the chiral condensate in the chiral limit 
(and at $\theta=0$). 
We will in fact always take this to be diagonal of entries 
$\mu_i = m_i\Sigma V$. In this
section we shall restrict ourselves 
to $N_f$ (physical) fermions with no additional quark species, and $N_v$
will thus here be taken to be zero from the very beginning.
The integration in eq. (\ref{ZUE}) is over the
coset of chiral symmetry breaking, here extended from SU($N_f$) to 
U($N_f$) due to the projection on a sector of fixed topological 
charge $\nu$. The integral is known in closed form \cite{JSV},
\beq
{\cal Z}_\nu^{(N_f)} (\{\mu\}) ~=~ \frac{\det A(\{\mu\})}{\Delta(\{\mu^2\})}
\label{ZUEclosed}
\eeq
where the $N_f\times N_f$ matrix $A$ is given by
\beq
A(\{\mu\})_{ij}=\mu_i^{j-1}I_{\nu+j-1}(\mu_i) ~,
\label{Adef}
\eeq
$I_n(x)$ is a modified Bessel function, and the denominator 
is given by the Vandermonde determinant of rescaled masses:
\beq
\Delta (\{\mu^2\})\ \equiv\ \prod_{i>j}^{N_f}(\mu_i^2-\mu_j^2)
\ =\ \det_{i,j}\left[ (\mu_i^2)^{j-1}\right] \ . 
\label{Vandermonde}
\eeq

\noi
Although the effective partition function is known explicitly, eq. 
(\ref{ZUEclosed}) is not in a form suitable for the derivation of spectral 
sum rules of the Dirac operator. To get it into such a form, it is
convenient to start with the case $\nu=0$. In that case the left and right
invariance of the Haar measure under unitary transformations shows that
the partition function depends only on the combination ${\cal M}^{\dagger}
{\cal M}$. A crucial observation of ref. \cite{GN,MMS} is that the partition
function in fact satisfies an infinite set of constraint equations that
can be used to determine it uniquely. The precise form of these constraints
depend on the chosen variables, and one can use two different sets:
\beq
t_k^+ ~\equiv~ \frac{1}{4^k k}{\rm Tr}(\m2)^{k} ~,  \  \  \ k=1,2,\ldots ~, 
\label{tk+}
\eeq
which is a suitable set for small-mass expansions, and 
\beq
t_k^- ~\equiv~ -\frac{2^{2k+1}}{2k+1}{\rm Tr}
\left( (\m2)^{-(2k+1)/2} \right) ~ , \ \ \  k=0,1,\ldots ~,
\label{tk-}
\eeq
which is a suitable set for large-mass expansions (the coefficients in front
have been inserted for later convenience). Spectral sum rules of 
the Dirac operator are derived by means of a small-mass expansion, and we 
shall therefore begin with the description in terms of the  
variables\footnote{Since $\m2$ is a hermitian matrix of size $N_f\times N_f$, 
only  
$N_f$ of these variables are independent; this is of no consequence for
the subsequent analysis.} $t_k^+$. 
\vspace{3mm}

\noi
Defining, for $n \geq 1$,  
\beq 
{\cal L}^+_n ~\equiv~ N_f\frac{\partial}{\partial t^+_{n}} +
\sum_{k=0}^{\infty} kt^+_{k}\frac{\partial}{\partial t^+_{n+k}} +
\sum_{k=1}^{n-1}\frac{
\partial^2}{\partial t^+_{k}\partial t^+_{n-k}} ~,
\eeq
the partition function (\ref{ZUE+}) is found to satisfy \cite{MMS}
\beq
{\cal L}^+_n {\cal Z}_{0}^{(N_f)} ~=~ \delta_{n,1} {\cal Z}_{0}^{(N_f)} ~.
\label{L+}
\eeq
These constraints are consistent in the sense that
they satisfy the classical Virasoro algebra
\beq
[{\cal L}^+_n, {\cal L}^+_m] ~=~ (n-m) {\cal L}^+_{n+m} ~,\label{Virasoro}
\eeq
so that they do not generate new constraints beyond the infinite tower
of ${\cal L}^+_n$'s. The constraints are also complete: They determine
the partition function uniquely, given the boundary condition that
${\cal Z}_{0}^{(N_f)}=1$ for all $t_k^+=0$. 

\noi
The small-mass power series for ${\cal Z}_{0}^{(N_{f})}$ can conveniently
be chosen \cite{MMS}
\beq
{\cal Z}_{0}^{(N_{f})} = 1 + \sum_M\sum_{1\leq k_1\ldots\leq k_M}
C_{N_f}(\{k\})\frac{ k_1t^+_1 \cdots k_Mt^+_M}{(k_1+\ldots+k_M)!} ~,
\label{ZUE+}
\eeq
where use has been made of the boundary condition to determine the zeroth 
order coefficient.

\noi
The coefficients $C_{N_f}(\{k\})$ are determined uniquely by the Virasoro
constraints (\ref{L+}). They factorize into a polynomial
part,  $\hat{C}_{N_f}(\{k\})$, and a singular part \cite{MMS}:
\beq
C_{N_f}(\{k\})~\equiv~\hat{C}_{N_f}(\{k\})\prod_{l=0}^{K(\{k\})-1}
(N_f^2-l^2)^{-1} ~ ,
\eeq
where $K(\{k\})\equiv\sum_i k_i$. The singular parts will play an
important r\^{o}le in what follows. Originally noted by de Wit and `t Hooft
in the context of strong-coupling expansions in lattice gauge theory
\cite{Dt}, these poles are in fact innocuous as far as the expansion
of the partition function itself is concerned \cite{B}. What happens is
that for (integer and non-vanishing!) values of $N$ the poles only occur
in terms that are linearly dependent, and therefore should be combined. After
combining them, all poles cancel. This phenomenon becomes obvious when
we consider the explicit expansion below.

\noi
The coefficients $\hat{C}_{N_f}(\{k\})$ have already been computed
recursively in ref. \cite{MMS} to high order, enough to determine the partition
function to $8^{\rm th}$ order in the masses. The restriction to the $\nu=0$
sector can easily be lifted due to {\sl flavor-topology duality}, which
states that  
\beq
{\cal Z}_{\nu}^{(N_f)}({\cal M},{\cal M}^\dagger) ~=~ [\det({\cal 
M})]^{\nu}{\cal  Z}_{0}^{(N_f+\nu)}({\cal M\cal M}^\dagger)
\eeq
where  ${\cal Z}_{0}^{(N_f+\nu)}({\cal M\cal M}^\dagger)$
is the $\nu=0$ partition
function extended with $\nu$ massless quarks \cite{Jac}.
Introducing $N\equiv N_f+N_v+\nu$ (here with $N_v=0$) we have 
\beqn
{\cal Z}_{\nu}^{(N_f)} & = & {\rm det}^\nu({\cal  M})\left( 1 +
\frac{1}{4N}\Tr\m2+\frac{1}{32(N^2-1)}(\Tr(\m2))^2-
\frac{1}{32N(N^2-1)}\Tr(\m2)^2 \right. \nn \\ & &
+\frac{1}{96N(N^2-1)(N^2-4)}\Tr((\m2)^3)-\frac{1}{128(N^2-1)(N^2-4)}\Tr(\m2)\Tr(
(\m2)^2) 
\nn \\ & &
+\frac{N^2-2}{384N(N^2-1)(N^2-4)}(\Tr(\m2))^3-\frac{5}{1024N(N^2-1)(N^2-4)(N^2-9
)}\Tr((\m2)^4)
\nn \\ & &
+\frac{2N^2-3}{768N^2(N^2-1)(N^2-4)(N^2-9)}\Tr(\m2)\Tr((\m2)^3)
\nn \\ & &
+\frac{N^2+6}{2048N^2(N^2-1)(N^2-4)(N^2-9)}(\Tr(\m2)^2)^2
\nn \\ & &
-\frac{1}{1024N(N^2-1)(N^2-9)}(\Tr(\m2))^2\Tr((\m2^2))
\nn \\ & &\left.
+\frac{N^4-8N^2+6}{6144N^2(N^2-1)(N^2-4)(N^2-9)}(\Tr(\m2))^4+\ldots \right)
\label{ZUEexp+}
\eeqn
We have checked that this expansion is correct to the order
given by expanding, for fixed $N_f$, the closed expression (\ref{ZUEclosed})
up to that order. (Note that it is not at all simple to rearrange the
closed expression (\ref{ZUEclosed}) in terms of an expansion in just
powers of traces of the mass matrix; the rearrangements are different for
each value of $N_f$). In this expansion we directly verify the cancellation
of de Wit-'t Hooft poles for any of the finite, integer, values involved.

\noi
With this high-order expansion of the effective partition function, we
can now easily derive a long series of spectral sum rules, following the
method of Leutwyler and Smilga \cite{LS}. 
Because the method is described in detail in ref. \cite{LS}, we shall not
give details of how these sum rules are extracted, but only quote the
results. We rescale as usual the eigenvalues $\lambda_i$ of the Dirac operator
 according to $\zeta_i \equiv \lambda_i \Sigma V$. Expanding the original
partition function (\ref{Zoriginal}) around the massless theory, and
comparing with the above expansion up to $8^{\rm th}$ order, we then find:
\beqn
\left\langle\sum_{\zeta_n>0}\frac{1}{\zeta_n^2}\right\rangle & = & 
\frac{1}{4}\frac{1}{N} \label{sum1}\\
\left\langle\sum_{\zeta_n>0}\frac{1}{\zeta_n^4}\right\rangle & = & 
\frac{1}{16}\frac{1}{N(N^2-1)} \label{sum2}\\
\left\langle\sum_{\zeta_n>0}\frac{1}{\zeta_n^6}\right\rangle & = &
\frac{1}{32}\frac{1}{N(N^2-1)(N^2-4)} \label{sum3}\\
\left\langle\sum_{\zeta_n>0}\frac{1}{\zeta_n^8}\right\rangle & = &
\frac{5}{256}\frac{1}{N(N^2-1)(N^2-4)(N^2-9)} ~ . \label{sum4}
\eeqn

\beqn
\left\langle\left(\sum_{\zeta_n>0}\frac{1}{\zeta_n^2}\right)^2\right\rangle & =
&
\frac{1}{16}\frac{1}{N^2-1} \label{sum5}\\
\left\langle\left(\sum_{\zeta_n>0}\frac{1}{\zeta_n^2}\right)^3\right\rangle & =
&
\frac{1}{64}\frac{N^2-2}{N(N^2-1)(N^2-4)} \label{sum6}\\
\left\langle\left(\sum_{\zeta_n>0}\frac{1}{\zeta_n^4}\right)^2\right\rangle & =
&
\frac{1}{256}\frac{N^2+6}{N^2(N^2-1)(N^2-4)(N^2-9)} \label{sum7}
\eeqn

\beqn
\left\langle \sum_{\zeta_m, \zeta_n>0}\frac{1}{\zeta_m^2\zeta_n^4}
\right\rangle & = &
\frac{1}{64}\frac{1}{(N^2-1)(N^2-4)} \label{sum8}\\
\left\langle \sum_{\zeta_m, \zeta_n>0}\frac{1}{\zeta_m^2\zeta_n^6}
\right\rangle & = &
\frac{1}{256}\frac{2N^2-3}{N^2(N^2-1)(N^2-4)(N^2-9)} \label{sum9}\\
\left\langle \sum_{\zeta_k, \zeta_m,
\zeta_n>0}\frac{1}{\zeta_k^2\zeta_m^2\zeta_n^4} 
\right\rangle & = &
\frac{1}{256}\frac{1}{N(N^2-1)(N^2-9)} \label{sum10}\\
\left\langle \sum_{\zeta_k, \zeta_l, \zeta_m,
  \zeta_n>0}\frac{1}{\zeta_k^2\zeta_l^2\zeta_m^2\zeta_n^2} \right\rangle & = &
\frac{1}{256}\frac{N^4-8N^2+6}{N^2(N^2-1)(N^2-4)(N^2-9)} ~ .\label{sumlast}
\eeqn
Of these, the relations (\ref{sum1}), (\ref{sum2}) and (\ref{sum5}) 
have been derived
previously \cite{LS} (see also ref. \cite{V0}). We note that the 
DeWit--`t Hooft poles make these sum rules divergent for an increasingly
larger number of integer $N$-values. This is due to infrared divergences 
near $\zeta \sim 0$. Just as these poles cancel among different terms
in the small-mass expansion, one can form analogous combinations of
sum rules that are infrared finite. For example, the de Wit-'t Hooft
poles at $N=1$ associated with the terms of order $\mu^4$ in eq. 
(\ref{ZUEexp+}) cancel by combining the third and fourth terms in that
expansion. This directly translates into a cancellation between two
eigenvalue sum rules from our list:
\beq
\left\langle\left(\sum_{\zeta_n>0}\frac{1}{\zeta_n^2}\right)^2\right\rangle
- \left\langle\sum_{\zeta_n>0}\frac{1}{\zeta_n^4}\right\rangle ~=~
\frac{1}{16}\frac{1}{N(N+1)} ~, 
\eeq
a cancellation that was already noticed by Leutwyler and Smilga \cite{LS}.
Similarly, while sum rules (\ref{sum3}), (\ref{sum6}) and (\ref{sum8}) 
are individually singular at both $N=1$ and $N=2$, the combination
\beq
2\left\langle\sum_{\zeta_n>0}\frac{1}{\zeta_n^6}\right\rangle + 
\left\langle\left(\sum_{\zeta_n>0}\frac{1}{\zeta_n^2}\right)^3
\right\rangle
- 3\left\langle \sum_{\zeta_m, \zeta_n>0}\frac{1}{\zeta_m^2\zeta_n^4}
\right\rangle
~=~ \frac{1}{64}\frac{1}{N(N+1)(N+2)}
\eeq
is finite (this is 6 times the ${\cal O}(1/\la^6)$ contribution to the
partition function). A similar phenomenon persists to all orders.

\noi
The massless spectral sum rules (\ref{sum1})-(\ref{sumlast}) can of course
be compared with the chiral Random Matrix Theory result. This is done by 
averaging the inverse moments according to the universal 
massless microscopic spectral density \cite{V}
\beq
\rho_S^{(N_{f},\nu)}(\zeta) ~=~ \frac{|\zeta|}{2}\left[J_{N_{f}+\nu}(\zeta)^2
- J_{N_{f}+\nu-1}(\zeta)J_{N_{f}+\nu+1}(\zeta)\right] ~,\label{rhomassless}
\eeq
$e.g.$,
\beq
\left\langle\sum_{\zeta_n>0} \frac{1}{\zeta_n^{2k}}\right\rangle ~=~ 
\int_{-\infty}^{\infty}\!d\zeta ~ \frac{\rho_S(\zeta)}{\zeta^{2k}} ~.
\eeq
We have explicitly checked a number of the above sum rules, always finding
completely agreement with the Random Matrix Theory results.

\section{The partially quenched chiral condensate}

\noi
Let us now extend the effective theory with $N_v$ valence quark fields. We
are interested in obtaining the partially quenched chiral condensate, not 
higher order susceptibilities, and therefore take the valence quark masses as
degenerate and of size $\mu_v=m_v\Sigma V$. As mentioned above,
the extended partition function is also the
generating functional for the $n$-point functions of $\bar{\psi}\psi$
(quenched or unquenched); the 
partially quenched chiral condensate is obtained through the replica method.
Since the partition function for $\nu=0$ only depends on $\m2$ and the valence
quarks
have degenerate masses it is natural to introduce a mass
matrix of the physical quarks $M$ so that 
\beq
\Tr(\m2) ~ = ~  \Tr(\mb2) + N_v \mu_v^2  ~ .\label{m2Nv+}
\eeq 
In short, for the definition of the partially quenched 
chiral condensate we can take
\beq
\frac{\Sigma_\nu(\mu_v,\{\mu\})}{\Sigma} ~\equiv~
\lim_{N_v\to0}\frac{1}{N_v}\frac{\partial}
{\partial\mu_v}{\rm ln} {\cal Z}_{\nu}^{(N_f+N_v)} ~ . 
\label{pqSigma}
\eeq
The subscript on $\Sigma$ refers to the topological sector, as usual.
As we have seen above, it is not particularly helpful to know the exact
analytical expression for the (extended) effective partition function 
${\cal Z}_{\nu}^{(N_f+N_v)}$ if one wishes to derive spectral sum rules
for the Dirac operator. There a small-mass expansion is required by definition.
As for finding the partially quenched chiral condensate using the
expression (\ref{pqSigma}), no expansion in either large or small masses
is required in principle. However, the closed analytical formula
(\ref{ZUEclosed}) is unfortunately not directly suitable from our point
of view. This is because the $N_v$-dependence enters in a quite non-trivial
way through the size of the matrix whose determinant needs to be taken.
Because there is no simple extension of the analytical expression
(\ref{ZUEclosed}) outside integer values of $N_v$, we restrict
ourselves here to series expansions. We shall in fact be able to carry such
expansions through in both the large-mass and small-mass regions.

\subsection{The small-mass expansion}

We begin with the small-mass expansion, because here we already have the
needed expansion at hand (see eq. (\ref{ZUEexp+})). Our first observation
is that $N_v$ enters in a manner that allows for an analytic continuation
once we make use of eq. (\ref{m2Nv+}). We can thus proceed with the
replica method.

\noi
Inserting the expansion (\ref{ZUEexp+}) into (\ref{pqSigma}) we obtain the
partially quenched chiral condensate to $7^{\rm
  th}$ order in the masses (truncating the expansion at this order is
a matter of choice; we do it because it corresponds to consistently
expanding both in the valence quark mass $\mu_v$ and the physical fermion
masses $\mu_i$ to the same order): 
\beqn
\frac{\Sigma_\nu(\mu_v,\{\mu\})}{\Sigma} & = & \frac{\nu}{\mu_v}+\mu_v
\left[\frac{1}{2 N}
+\frac{{\rm Tr}\mb2 }{8N^2(N^2-1)}
+\frac{({\rm Tr}\mb2)^2}{8N^3(N^2-1)(N^2-4)}\right. 
\nn \\  & & 
 -\frac{{\rm Tr}M^4}{16N^2(N^2-1)(N^2-4)} +
  \frac{15{\rm Tr}M^6}{384N^2(N^2-1)(N^2-4)(N^2-9)}
\nn \\  & &  
\left.+\frac{3(2N^2-3)({\rm
    Tr}\mb2)^3}{32N^4(N^2-1)^2(N^2-4)(N^2-9)}
+\frac{3(3-2N^2){\rm Tr}\mb2{\rm
Tr}M^4}{32N^3(N^2-1)^2(N^2-4)(N^2-9)} + \ldots \right] 
\nn \\  & &  
-\mu_v^3\left[
   \frac{1}{8N(N^2-1)}+\frac{{\rm Tr}\mb2}{8N^2(N^2-1)(N^2-4)}\right. 
\nn \\  & & 
 \left.+\frac{(6N^2-9)({\rm Tr}\mb2)^2}{32N^3(N^2-1)^2(N^2-4)(N^2-9)}\right.
\nn \\
& & -\left.\frac{3(3N^2-7){\rm Tr}M^4}{128N^2(N^2-1)^2(N^2-4)(N^2-9)}
+ \ldots\right] \nn \\   & &
 +\mu_v^5\left[\frac{1}{16N(N^2-1)(N^2-4)}+\frac{15{\rm Tr}\mb2}
{128N^2(N^2-1)(N^2-4)(N^2-9)} + \ldots \right] 
\nn    \\ & &
 -\mu_v^7\frac{5}{128N(N^2-1)(N^2-4)(N^2-9)}+\ldots
\label{s-mpqcc}
\eeqn
where $N=N_f+\nu$ (since the $N_v\to 0$ limit has already been taken).
Note that the ``topological'' term $\nu/\mu_v$ comes out trivially
from the $(\det {\mathcal M})^{\nu}$-factor in front, as it should.

\noi
Is this expansion correct? There are by now two independent ways to check it:
Comparing with Random Matrix Theory, and comparing with the supersymmetric
method of partially quenched Lagrangians. Before proceeding with such
comparisons, let us however first note an obvious feature of the
above expansion. In contrast to the small-mass expansion of the
partition function itself (\ref{ZUEexp+}), there is clearly {\em not}
going to be a cancellation of the de Wit-`t Hooft poles in the expansion
of the partially quenched chiral condensate. This may appear surprising,
since the chiral condensate is just determined from the partition
function through, for example, a formula like (\ref{firstdef}). If the
de Wit-`t Hooft poles cancel in the partition function, how can they not
cancel after taking the $\mu_v$-derivative and the limit $N_v\to 0$ as
in eq. (\ref{firstdef})? The answer is simple. To
be able to take the quenched limit of $N_v \to 0$ we must not cross any 
singularities that can obstruct the analytical continuation. This is,
however, precisely what happens for those terms in the expansion 
(\ref{s-mpqcc}) that, for given $N$, are singular. To give an example,
let us consider the $\mu_v^3$-term in the expansion (\ref{s-mpqcc}). For 
simplicity, let us also consider the case
where all physical masses vanish. Then there is only one term: 
\beq
\frac{\mu_v^3}{8N(N^2-1)} ~. \label{mu^3}
\eeq
This term is singular for $N=1$ (and, of course, also in the more trivial 
case of $N=0$), and therefore cannot possibly represent the partially
quenched chiral condensate to order $\mu_v^3$ in the $N=1$ theory. Let us
therefore trace what happens if we set $N_f+\nu=1$ from the outset. The
only term in the expansion (\ref{ZUEexp+}) that gives an ${\cal O}(\mu_v^3)$
contribution to $\Sigma_{\nu}(\mu_v)$ in this massless case is ($N\equiv
N_f+N_v+\nu$)
\beqn
\frac{1}{32N(N^2-1)}\Tr(\m2)^2 &=& \frac{N_v\mu_v^4}{32N(N^2-1)}\cr
&=& \frac{\mu_v^4}{32(N_v+1)(N_v+2)} ~,
\eeqn
which precisely in this $N_f+\nu=1$ case is not linear in $N_v$. The
factor $1/N_v$ in the expression for the quenched condensate (\ref{pqSigma})
will in this case remain uncancelled, and an analytic continuation
to $N_v=0$ is prohibited by the singularity. The term (\ref{mu^3})
in the expansion (\ref{s-mpqcc}) is therefore incorrect precisely in
the case $N_f+\nu=1$ (but it is valid for all $N_f+\nu > 1$). 
This holds also for all higher orders in the expansion. The small-mass
expansion for $\Sigma_{\nu}$ is thus given to a finite number of terms.
For example, in the {\em massless} $N_f+\nu=3$ theory we obtain
\beq
\frac{\Sigma_\nu(\mu_v,\{\mu\})}{\Sigma} ~=~ \frac{\nu}{\mu_v} + \frac{\mu_v}{6}
- \frac{\mu_v^3}{192} + \frac{\mu_v^5}{1920} + ...~, ~~~~~~~~~~~~
N_f+\nu=3 ~,
\eeq
where the higher-order terms in $\mu_v$ cannot be purely power-like.
The same phenomenon occurs for larger integer values of $N$, 
where we encounter the singularity at higher orders. We thus
conclude that the small-mass expansion (\ref{s-mpqcc}) is valid 
up to the first term where a pole appears. The obstruction to an
all-order expansion of $\Sigma_{\nu}(\mu_v,\{\mu_i\})$ is
from this point of view
uncancelled de Wit-`t Hooft poles. The expansion can of course
be pushed to very high orders by considering $N_f+\nu$ sufficiently large.
We stress again that the expansion is perfectly meaningful up to the
order at which the first uncancelled de Wit-`t Hooft pole appears, and
that it is completely understood why the expansion cannot be pushed beyond
this order. It is not a peculiar and unwanted artifact of the replica 
method, but a real feature of the small-mass expansion of the partially
quenched chiral condensate.

\noi
With this caveat in mind, we can now check the expansion
(\ref{s-mpqcc}) against both the supersymmetric method and results from
Random Matrix Theory. For instance, both methods yield a fully quenched
chiral condensate of the form \cite{VMC,OTV}
\beq
\frac{\Sigma_{\nu}(\mu_v)}{\Sigma} ~=~ \mu_v\left[I_{\nu}(\mu_v)
K_{\nu}(\mu_v) + I_{\nu+1}(\mu_v)K_{\nu-1}(\mu_v)\right] + 
\frac{\nu}{\mu_{v}} ~. \label{sigmaNf0}
\eeq
where $I_n(x)$ and $K_n(x)$ are modified Bessel functions.
Looking at the small-$\mu_v$ expansions of $K_n(\mu_v)$, one notices
that they are only purely power-like up to order $n-2$ (after which a
logarithmic term must be included). The
small-mass expansion of the closed expression (\ref{sigmaNf0}) therefore
cannot possibly match to all orders the purely power-like expansion 
that would seem to follow from the small-mass expansion of the partition 
function in eq. (\ref{ZUEexp+}).
And we find that the agreement
of the two expansions is {\em exact} precisely up to the order at
which our small-expansion (\ref{s-mpqcc}) ceases to be valid. The
appearance, at higher orders, of logarithmic terms in the expansion
explicitly confirms the statement above: there is simply no power-law
small-mass expansion of the chiral condensate beyond that given order.
It is quite remarkable that the normally cancelling de Wit-`t Hooft
poles precisely obstruct the small-mass expansion at just the right order.

\noi
We have also checked that the double-expansions in both $\mu_v$ and
physical fermion masses $\mu_i$ are correct precisely up to the order
at which eq. (\ref{s-mpqcc}) is valid. We have done this by comparing
the small-mass expansions of the partially quenched chiral condensate in
the $N_f=1$ theory with the expansions of the closed expression
\beqn
\frac{\Sigma_{\nu}(\mu_v,\mu)}{\Sigma} &=& \mu_v\left[I_{\nu+1}(\mu_v)
K_{\nu+1}(\mu_v) + I_{\nu+2}(\mu_v)K_{\nu}(\mu_v)\right] + 
\frac{\nu}{\mu_{v}} \cr
&& + 2\mu\frac{K_{\nu}(\mu_v)}{I_{\nu}(\mu)}\frac{\mu_vI_{\nu}(\mu_v)
I_{\nu+1}(\mu)-\mu I_{\nu}(\mu)I_{\nu+1}(\mu_v)}{\mu_v^2-\mu^2} ~,
\eeqn
which has been derived by the supersymmetric method \cite{OTV} (and which
can also easily been seen to agree with the Random Matrix Theory result).
Again we find perfect agreement with all terms up to the order at which
our expansion (\ref{s-mpqcc}) ceases to be valid, $i.e.$ up to the first 
uncancelled de Wit-`t Hooft pole.

\subsection{The large-mass expansion}

Having succeeded in deriving the low-mass expansion for the partially
quenched chiral condensate, we next turn to the opposite expansion,
$i.e.$ for large masses. The advantage of this expansion is that it
can be pushed to arbitrarily high order, for any value of $N_f$.

\noi
Recall that the effective partition function (\ref{ZUE}) only depends on the
combination
$\m2$ for $\nu=0$. This combination is hermitian and there are thus only $N_f$
degrees of freedom in ${\cal Z}_{0}^{(N_f)}$, represented for instance
by the eigenvalues $\mu_i^2$ of $\m2$.
In terms of $\mu_i$ the expansion variables of eq. (\ref{tk-}) read
\beq
t_k^- ~ \equiv~ -\frac{2^{2k+1}}{2k+1}\sum_{i=1}^{N_f} \mu_i^{-(2k+1)}  ~ .
\label{tk-(mu)}
\eeq

\noi
The idea is now to find suitable partition function constraints that will
enable us to solve for the partition function in an expansion in the
variables $t_k^-$. This is not completely straightforward, but as shown
first by Gross and Newman \cite{GN}, it is possible to recover a
complete set of Virasoro constraints in the variables $t_k^-$ if one first
extracts a simple prefactor from the partition function. The factorization
is as follows \cite{GN} (see also ref. \cite{MMS} for the generalization
to other types of matrix integrals):
\beq
{\cal Z}_{0}^{(N_f)}  ~=~  (\prod_{a,b}^{N_f}(\frac{1}{2}\mu_a+
\frac{1}{2}\mu_b)^{-1/2})e^{\sum_b
  \mu_b}~ Y_0^{(N_f)}(\{t_k^-\})  ~ \label{ZUE-} ~ .
\eeq
The exponential prefactor indicates that an expansion of the partition
function entirely in terms of the $t_k^-$'s simply is not possible. This
is also obvious if we consider the original group integral in a saddle-point
approximation around large ${\cal M}$: the leading behaviour is indeed
exponentially large.
Moreover, one suspects from the form (\ref{ZUE-}) which is essentially
a saddle-point expansion that
the resulting series expansion we will derive below
is at best asymptotic. This is in contrast to the previously considered
small-mass expansion, which we believe is convergent. 

\noi
The remaining factor $Y_0^{(N_f)}(\{t_k^-\})$ in eq. (\ref{ZUE-})
turns out to be annihilated by 
an infinite set of Virasoro operators. Explicitly, by defining  
\beqn
{\cal L}^-_0 & \equiv &
\sum_{k=0}^\infty(k+\frac{1}{2})t^-_k\frac{\partial}{\partial
  t^-_k}+\frac{1}{16}+\frac{\partial}{\partial t^-_0}  ~ ,\label{L0-}  \\
 {\cal L}^-_n & \equiv &
\sum_{k=0}^\infty(k+\frac{1}{2})t^-_k\frac{\partial}{\partial
  t^-_{k+n}}+\frac{1}{4}\sum_{k=1}^{n}\frac{\partial^2}{\partial
  t^-_{k-1}\partial t^-_{n-k}}+\frac{\partial}{\partial
  t^-_n}   ~, \ \ \  n\geq1 
\label{Ln-}
\eeqn
the function $Y_0^{(N_f)}(\{t_k^-\})$ 
of (\ref{ZUE-}) is found to satisfy \cite{GN}
\beq
{\cal L}^-_n Y_0^{(N_f)}  ~=~ 0 \ \ \ n\geq 0 ~ .
\label{L-}
\eeq
One readily verifies that these constraints indeed also fulfill the
commutation relations of classical Virasoro generators (\ref{Virasoro}),
and thus form a consistent set of constraints.
These constraints are also complete in that they determine
the partition function uniquely, given the boundary condition
$Y_0^{(N_f)}=1$ for $t^-_k=0$, $i.e.$ in the limit of infinite masses. 
While these constraint equations were already established in 
ref. \cite{GN}, they
have not previously been used to derive a systematic expansion for
the partition function. We do this by copying the procedure of the
small-mass expansion. That is, we expand the unknown function 
$Y_0^{(N_f)}(\{t_k^-\})$ as follows:
\beq 
Y_0^{(N_f)}(\{t_k^-\})  ~ \equiv ~  1 + \sum_{k=0}^\infty c_k t^-_k +
\sum_{0\leq k_1\leq k_2} c_{k_1,k_2} t^-_{k1} t^-_{k_2} + \ldots
\label{Y}
\eeq

\noi
We next solve for the coefficients  $c_{k_1,\ldots,k_n}$ iteratively. In fact,
this is in many ways simpler than in the small-mass expansion in that
we in this case easily can derive closed analytical expressions to
all orders. For example, we find the following simple formula for
the first strings of coefficients: 
\beqn
c_{ {0,0,\ldots,0,k,\ldots,k \atop n+1 \ \ \ m }}
=\left(\prod_{j=1}^n\frac{-1}{j+1}
\frac{8m(2k+1)+8j+1}{16}\right)c_{{0,k,\ldots,k, \atop \ \ m }} 
~ .
\eeqn

\noi
Using this formula, and others similar,
it is straightforward to derive very high order
expansions for $Y_0^{(N_f)}(\{t_k^-\})$, and therefore also for the partition
function itself. For example, to $7^{\rm th}$ order in the masses we find
a rather formidable-looking expression (that is easily pushed to much higher
orders): 
\beqn
{\cal Z}_{0}^{(N_f)} & = &
(\prod_{a,b}^{N_f}(\frac{1}{2}\mu_a+\frac{1}{2}\mu_b)^{-1/2})
\exp\left[\sum_b
  \mu_b\right] 
\nn \\  & & 
\left[ 1+\frac{1}{8}\Tr(\m2)^{-1/2}+\frac{9}{128}(\Tr(\m2)^{-1/2})^2
 \right. 
\nn \\  & & 
+\frac{51}{1024}(\Tr(\m2)^{-1/2})^3
+\frac{3}{128}\Tr(\m2)^{-3/2}
+\frac{1275}{32768}(\Tr(\m2)^{-1/2})^4
\nn \\  & & 
+\frac{75}{1024}(\Tr(\m2)^{-1/2})(\Tr(\m2)^{-3/2})
+\frac{8415}{262144}(\Tr(\m2)^{-1/2})^5
\nn \\  & &
+\frac{2475}{16384}(\Tr(\m2)^{-1/2})^2(\Tr(\m2)^{-3/2})
+\frac{45}{1024}\Tr(\m2)^{-5/2}
+\frac{115005}{4194304}(\Tr(\m2)^{-1/2})^6
\nn \\  & &
+\frac{33825}{131072}(\Tr(\m2)^{-1/2})^3(\Tr(\m2)^{-3/2})
+\frac{6075}{98304}(\Tr(\m2)^{-3/2})^2
\nn \\  & &
+\frac{1845}{8192}(\Tr(\m2)^{-1/2})(\Tr(\m2)^{-5/2})
+\frac{805035}{33554432}(\Tr(\m2)^{-1/2})^7
\nn \\  & &
+\frac{1657425}{4194304}(\Tr(\m2)^{-1/2})^4(\Tr(\m2)^{-3/2})
+\frac{99225}{262144}(\Tr(\m2)^{-1/2})(\Tr(\m2)^{-3/2})^2
\nn \\  & &
\left.+\frac{90405}{131072}(\Tr(\m2)^{-1/2})^2(\Tr(\m2)^{-5/2})
+\frac{7875}{32768}\Tr(\m2)^{-7/2} + \ldots
  \right] 
\label{l-mpqcc}
\eeqn

\noi
In writing (\ref{tk-}) and (\ref{tk-(mu)}) we have implicitly assumed 
that $\m2$ is
invertible. This means that there must be no zero-eigenvalues in the mass
matrix, and we thus by default cannot consider any number of massless fermions
in this large-mass expansion. In particular, the concept of flavor-topology 
duality, by which a topological charge $\nu$ is implemented by adding
$\nu$ massless fermions to the $\nu=0$ partition function, 
cannot be applied in the framework of a large-mass expansion. 
Because the Virasoro constraints (\ref{L0-}) and (\ref{Ln-}) apply only
to the case $\nu=0$ we know of no analogous way to implement this large-mass
expansion outside of the $\nu=0$ sector.

\noi 
Introducing $N_v$ mass-degenerate valence quarks into the large-mass
expansion of the effective partition function, and noting that in this
case
\beq
\Tr(\m2)^{-1/2} ~=~ \Tr(M)^{-1} + N_v \mu^{-1} ~,
\eeq
we again observe that $N_v$ can 
be regarded as a continuous parameter. The replica approach can therefore 
again be applied.

\noi
Using the definition (\ref{pqSigma}) and a good handful of simple algebra 
we find the partially quenched chiral condensate 
for large masses:
\beqn
\frac{\Sigma_0(\mu_v,\{\mu\})}{\Sigma}  & = & 1 -
\sum_{i=1}^{N_f}\frac{1}{\mu_i+\mu_v}
\nn \\ & &
-\frac{1}{\mu_v^2}\left[\frac{1}{8}+\frac{1}{8}\Tr(\frac{1}{M})
+\frac{1}{8}\left(\Tr(\frac{1}{M})\right)^2
+\frac{9}{128}\Tr(\frac{1}{M^3})\right.
\nn \\ & &
+\frac{1}{8}\left(\Tr(\frac{1}{M})\right)^3 
+\frac{1}{8}\left(\Tr(\frac{1}{M})\right)^4
+\frac{9}{32}\Tr(\frac{1}{M^3})\Tr(\frac{1}{M})
+\frac{1}{8}\left(\Tr(\frac{1}{M})\right)^5
\nn \\ & &
\left. 
+\frac{45}{64}\Tr(\frac{1}{M^3})\left(\Tr(\frac{1}{M})\right)^2
+\frac{225}{1024}\Tr(\frac{1}{M^5})
+\ldots\right]
\nn \\ & &
-\frac{1}{\mu_v^4}\left[\frac{9}{128}
+\frac{27}{128}\Tr(\frac{1}{M})
+\frac{27}{64}\left(\Tr(\frac{1}{M})\right)^2
+\frac{45}{64}\left(\Tr(\frac{1}{M})\right)^3
+\frac{189}{512}\Tr(\frac{1}{M^3})
+\ldots\right]
\nn \\ & &
-\frac{1}{\mu_v^6}\left[\frac{225}{1024}
+\frac{1125}{1024}\Tr(\frac{1}{M}) +\ldots\right
]
\eeqn
This expansion agrees with the asymptotic expansion of the analytical
expressions for $\Sigma(\mu_v,\{\mu\})$ found in \cite{OTV} for $N_f=0$ and
 $N_f=1$.

\section{Massive spectral sum rules}

The spectral sum rules presented in section \ref{masslessSR} concern
inverse moments of the Dirac eigenvalues averaged with respect to the
massless theory. 
These massless spectral sum rules are, however, only special cases of
massive spectral sum rules obtained by averaging with respect to the 
massive theory. In fact, massive spectral sum rules appear quite naturally in
lattice simulations. Such massive spectral sum rules are conventionally taken
\cite{SV,D0} to be of the form
$$
\int_0^{\infty}\!d\zeta~ \rho_S(\zeta;\{\mu_i\})\prod_j\frac{1}
{(\zeta^2+\mu_j^2)^{n_j}} ~,
$$
for given integer $n_j$'s. Here masses and Dirac operator eigenvalues
are treated on equal footing in the denominators. However, 
we could equally well define massive spectral sum rules by 
\beq
\left\langle\sum_{\zeta_j>0}\frac{1}{\zeta^{2n}_j}\right\rangle ~=~
\int_0^{\infty}\!d\zeta~ \frac{\rho_S(\zeta;\{\mu_i\})}{\zeta^{2n}} ~,
\eeq
which are just the usual inverse moments, but evaluated in the massive
theory. Here and below the brackets denote the average with respect to the 
massive and partially quenched theory in a topological gauge field sector of
charge $\nu$. In order to derive such massive spectral sum rules we
return to the small-mass expansion of the partially quenched condensate
(\ref{s-mpqcc}). 

\noi
Inserting the original partition function (\ref{Zoriginal}) into the
definition (\ref{pqSigma}) of the partially quenched chiral condensate one
has
\beq
\frac{\Sigma_\nu(\mu_v,\{\mu\})}{\Sigma} ~=~
2\mu_v\left\langle\sum_{\zeta_n>0}\frac{1}{\zeta^2_n+\mu^2_v}\right\rangle+
\frac{\nu}{\mu_v}~ .\label{sigmageneral}
\eeq
We can now find a useful general relation between the partially
quenched chiral condensate, and (massive) spectral sum rules.
The connection is simple: the sum rules of inverse moments $\zeta^{-2n}$
of the eigenvalues will be convergent up to a given value of $n$, which
we denote by $k$. Let us therefore expand the denominator
of eq. (\ref{sigmageneral}) in partial fractions up to this maximal
value $k$:
\beq
\frac{1}{\zeta^2_n+\mu^2_v} ~=~ \frac{1}{\zeta^2_n} -
\frac{\mu_v^2}{\zeta^4_n} + \ldots + 
\frac{(-1)^{k}\mu_v^{2k}}{\zeta_n^{2k}(\zeta^2_n+\mu^2_v)} 
\label{partialfrac}
\eeq 
which means that
\beqn
\frac{\Sigma_\nu(\mu_v,\{\mu\})}{\Sigma} & = &
  2\left\langle\sum_{\zeta_n>0}\frac{1}{\zeta^2_n}\right\rangle \mu_v 
-2\left\langle\sum_{\zeta_n>0}\frac{1}{\zeta^4_n}\right\rangle \mu_v^3+\ldots+
% \nn \\  & & 
\left\langle\sum_{\zeta_n>0}\frac{2(-1)^k\mu_v^{2k+1}}
{\zeta_n^{2k}(\zeta^2_n+\mu^2_v)}\right\rangle
+\frac{\nu}{\mu_v} ~.
\eeqn 
We can thus simply read off a whole string of massive spectral sum 
rules from the coefficients in the $\mu_v$-expansion of the partially
quenched chiral condensate (\ref{s-mpqcc}):
\beqn 
\left\langle\sum_{\zeta_n>0}\frac{1}{\zeta^2_n}\right\rangle & = & 
\frac{1}{4 N} + \frac{{\rm Tr}\mb2 }{16N^2(N^2-1)}
+ \frac{({\rm Tr}\mb2)^2}{16N^3(N^2-1)(N^2-4)} 
\nn \\  & & 
 -\frac{{\rm Tr}((\mb2)^2)}{32N^2(N^2-1)(N^2-4)} +
  \frac{15{\rm Tr}((\mb2)^3)}{768 N^2(N^2-1)(N^2-4)(N^2-9)}
\nn \\  & &  
+\frac{3(2N^2-3)({\rm
    Tr}\mb2)^3}{64N^4(N^2-1)^2(N^2-4)(N^2-9)}\nn \\
& & +\frac{3(3-2N^2){\rm Tr}\mb2{\rm
Tr}((\mb2)^2)}{64N^3(N^2-1)^2(N^2-4)(N^2-9)} + \dots
\\  
\left\langle\sum_{\zeta_n>0}\frac{1}{\zeta^4_n}\right\rangle
& = &  \frac{1}{16N(N^2-1)}+\frac{{\rm Tr}\mb2}{16N^2(N^2-1)(N^2-4)}
\nn \\  & & 
 +\frac{(6N^2-9)({\rm Tr}\mb2)^2}{64N^3(N^2-1)^2(N^2-4)(N^2-9)}\nn \\
& &   -\frac{3(3N^2-7){\rm Tr}((\mb2)^2)}{256N^2(N^2-1)^2(N^2-4)(N^2-9)}
+ \ldots \\
\left\langle\sum_{\zeta_n>0}\frac{1}{\zeta^6_n}\right\rangle  & = &
 \frac{1}{32N(N^2-1)(N^2-4)}+\frac{15{\rm Tr}\mb2}
{256N^2(N^2-1)(N^2-4)(N^2-9)} + \ldots 
\\ 
\left\langle\sum_{\zeta_n>0}\frac{1}{\zeta^8_n}\right\rangle
& = & \frac{5}{256N(N^2-1)(N^2-4)(N^2-9)}+\ldots
\eeqn
The resulting sum rules are here only given as a perturbative expansion
in the physical fermion masses, in contrast to the massless spectral sum
rules which are exact. One notices that trivially the ``diagonal'' massless
spectral sum rules (\ref{sum1})-(\ref{sum4}) are reproduced by taking all
$\mu_i=0$. As discussed already in section 2, the above expansions
are valid for $N$ larger than the biggest integer poles. This is
completely understandable from the present point of view, 
as it corresponds to performing the expansion
(\ref{partialfrac}) in partial fractions only up to the point where
the spectral sums of all terms still converge. The ``non-diagonal'' massive
sum rules can also be calculated within this framework. One simply needs to
break the mass-degeneracy of the valence quarks and consider general 
partially quenched
$n$-point correlators of $\bar{\psi}\psi$, as discussed in the introduction.

\noi
These new massive spectral sum rules can easily be compared with the 
predictions from Random Matrix Theory. What we need is to expand 
the microscopic spectral density $\rho_S^{(N_f,\nu)}(\zeta,\{\mu_i\})$ in
powers of Tr$M^2$, and then insert this expansion in
\beq
\frac{\Sigma_\nu(\mu_v,\{\mu_i\})}{\Sigma} = 2\mu_v\int_0^\infty d\zeta
\frac{\rho_S^{(N_f,\nu)}(\zeta,\{\mu\})}{\zeta^2+\mu_v^2}+\frac{\nu}{\mu_v} ~.
\label{sigmarhoconn}
\eeq
For example, for $N_f=1$ the microscopic spectral density for a massive
fermion in a gauge field sector of arbitrary topological index $\nu$ is
given by \cite{D0} (see also the second reference of \cite{ADMN} and ref.
\cite{WGW}):
\beqn
\rho_S^{(\nu)}(\zeta;\mu) &=& \frac{|\zeta|}{2}[J_{\nu+1}(\zeta)^2 -
J_{\nu}(\zeta)J_{\nu+2}(\zeta)] + \frac{|\zeta|\mu^2 J_{\nu}(\zeta)}{2(\nu+1)
(\zeta^2\! +\! \mu^2)}
\left[\frac{I_{\nu+2}(\mu)}{I_{\nu}(\mu)}
J_{\nu}(\zeta) + J_{\nu+2}(\zeta)\right] \cr
&=& \frac{|\zeta|}{2}[J_{\nu+1}(\zeta)^2 -
J_{\nu}(\zeta)J_{\nu+2}(\zeta)] + \frac{J_{\nu}(\zeta)J_{\nu+2}(\zeta)}
{2(\nu+1)|\zeta|}\mu^2 \cr
&& + \frac{1}{2(\nu+1)}\left(\frac{J_{\nu}(\zeta)}{4(\nu+1)(\nu+2)|\zeta|} -
\frac{J_{\nu+2}(\zeta)}{|\zeta|^3}\right)J_{\nu}(\zeta)\mu^4 + \ldots 
\eeqn 
Inserting this expansion into (\ref{sigmarhoconn}), we find that the
relevant integrals become analytically doable when $\nu$ is taken large
enough to make the integrals converge. In this way we have
explicitly confirmed a number of terms in the above expansions of the
massive spectral sum rules.

\setcounter{equation}{0}
\section{Generalizations and outlook}

\noi
The small and large mass expansions for the partially quenched chiral
condensate can of course in principle be extended to the two other
major universality classes, corresponding to SU(2) gauge group with
fermions in the fundamental representation, and SU($N_c\geq2$) with
fermions in the adjoint representation. The challenge is here to find
a convenient method that permits high-order expansions. In our present
case all simplifications arose from the observation that the effective 
partition function is annihilated by two sets of Virasoro constraints, 
which in turn is rooted in the fact that this effective partition function is
a $\tau$-function of an integrable KP hierarchy. The effective partition
functions of the two other universality classes have not been nearly
as well studied, although there are reasons to believe that they
are what is known as ``Pfaffian $\tau$-functions'' (see, $e.g.$, ref.
\cite{vdL}). It may thus be
possible to derive analogous partition function constraints, which will
permit high-order expansions with little labor. In this connection we
also make the following observations.
The  Virasoro constraints (\ref{L+}) and (\ref{L-}) both follow from the
simple differential equation
\beq
\frac{\partial^2 {\cal Z}_{0}^{(N_{f})}}{\partial {\cal M}_{ik}\partial
 {\cal  M}^\dagger_{kj}} ~ = ~ \delta_{ij}\frac{1}{4}{\cal Z}_{0}^{(N_{f})}
\label{partdiffeq}
\eeq
by changing variables to the pertinent expansion parameters. This 
differential equation by itself is a trivial consequence of the fact
that the integration manifold is U($N_f+N_v$), as it just amounts
to inserting the unitarity condition $U^{\dagger}U=1$ in the integrand
of the effective partition function (\ref{ZUE}). Imposing two boundary 
conditions is thus sufficient to determine the
partition function uniquely. In the case at hand it is the choice ${\cal 
  Z}_0^{(N_f)}(t^+_k=0)=1$ and $Y_0^{(N_f)}(t^-_k=0)=1$, together
with the required property
${\cal Z}_0^{(N_{f})}({\cal M},{\cal M}^\dagger)=
{\cal Z}_{0}^{(N_{f})}(\m2)$ that 
provide the two conditions.

\noi
Except for a change of the proportionality constant, here 1/4, the
differential equation (\ref{partdiffeq}) also holds for the effective partition
functions of the two other universality classes mentioned above, as well as
for the one of QCD$_3$. It is also instructive to notice that the restriction
to a sector of fixed topological charge does not affect the differential
equation. The boundary conditions imposed on the differential equation,
however, separate the solutions, $i.e.$, the partition functions. For
instance, before the projection onto a fixed topological sector the partition
function studied here depends on det$({\mathcal M})$ and det$({\mathcal
  M}^\dagger)$ as well as on ${\mathcal M}{\mathcal M}^\dagger$. 

\noi
Of course we can already at this stage obtain the partially quenched
chiral condensates to lowest non-trivial orders for the two remaining
ensembles by simply making use
of the same small-mass expansions of the partition function that were
used to derive the first sum rules. This is a trivial exercise: from
ref. \cite{SmV} we learn that the effective partition function
for SU(2) gauge group and $N_f+N_v$ fermions in the fundamental representation
is
\beq
{\cal Z}_{\nu} ~=~ [{\mbox{\rm Pf}}({\tilde{\mathcal M}})]^{\nu}\left(1 + 
\frac{1}{8(2N_f+ 2N_v+ \nu -1)}\Tr{\tilde{\mathcal M}}^{\dagger}
{\tilde{\mathcal M}}+ \ldots\right) ~,
\eeq
where, now ${\tilde{\mathcal M}}$ is an antisymmetric $(2N_f+2N_v)\times
(2N_f+2N_v)$ matrix with the usual mass matrices $+{\mathcal M}$ and
$-{\mathcal M}$ placed in the two off-diagonal blocks, and zeros in
the two diagonal blocks.
Using again eqs. (\ref{m2Nv+}) and (\ref{pqSigma}) we find the partially
quenched chiral condensate to this order:
\beq
\frac{\Sigma_{\nu}(\mu_v)}{\Sigma} ~=~ \frac{\nu}{\mu_{v}} +
\frac{\mu_{v}}{2(2N_f + \nu -1)} + \ldots ~,
\eeq
which one can confirm matches the first term in the expansion of the
result obtained from Random Matrix Theory (see the first of ref. \cite{DEHN}). 
Note that this expansion has its first uncancelled de Wit-`t Hooft pole
at $2N_f+\nu = 1$, and the above term in the expansion is thus valid
for $2N_f+\nu > 1$. 

\noi
Similarly, for the universality class corresponding to $N_f+N_v$
adjoint fermions and arbitrary gauge group SU($N_c$) we see from 
ref. \cite{SmV} that the partition function expansion is
\beq
{\cal Z}_{\bar{\nu}} ~=~ \det({\mathcal M})^{\bar{\nu}}\left(1 +
\frac{1}{2(N_f+N_v+2\bar{\nu} + 1)}\Tr\m2 + \ldots \right) ~,
\eeq
where $\bar{\nu}=\nu N_c$ is an integer. 
Using the present replica method this leads to a
partially quenched chiral condensate of   
\beq
\frac{\Sigma_{\bar{\nu}}(\mu_v)}{\Sigma} ~=~ \frac{\bar{\nu}}{\mu_{v}} +
\frac{\mu_v}{N_f+ 2\bar{\nu}+1} + \ldots ~,
\eeq
which one again can check matches the result from Random Matrix Theory
\cite{DEHN}. There is no de Wit-`t Hooft pole in this case, and
thus no retriction on the validity of this first term.

\noi
Since one can obtain perturbative solutions to the partially
quenched chiral condensate in all three universality classes, a natural
question concerns the microscopic spectral density of the Dirac
operator $\rho_S(\zeta;\{\mu_i\})$. This density is given by the
discontinuity of the partially quenched chiral condensate
across the cut on the imaginary axis \cite{OTV}:
\beq
\rho_S(\zeta;\{\mu_i\}) ~=~ \frac{1}{2\pi}
\left .{\rm Disc}\right |_{\mu_v = i\zeta}\Sigma_{\nu}(\mu_v,\{\mu_i\}) 
~=~ \frac{1}{2\pi}\lim_{\epsilon \rightarrow 0}
[\Sigma_{\nu}(i\zeta+\epsilon,\{\mu_i\}) - 
\Sigma_{\nu}(i\zeta-\epsilon,\{\mu_i\})] 
~ . \label{spectdisc}
\eeq
(This identification holds when one considers
$\Sigma_{\nu}(\mu_v,\{\mu_i\})$ as a function of a {\em real} mass
$\mu_v$, and then replaces $\mu_v \to i\zeta \pm\epsilon$). It now
seems straightforward to insert our small-mass and large-mass series
expansions for $\Sigma_{\nu}(\mu_v,\{\mu_i\})$, and derive corresponding
series expansions for the spectral density $\rho_S(\zeta;\{\mu_i\})$.
But this is not possible. Let us first consider the small-mass expansion
of eq. (\ref{s-mpqcc}). As we emphasized in section 3.1, this expansion
is valid up to the first de Wit-`t Hooft pole. Except for the topological
term $\nu/\mu_v$ which always trivially yields the correct $\delta$-function
contribution to the microscopic spectral density, this power-series
expansion of a finite number of terms has no cut across the imaginary axis. 
This may seem surprising, since the microscopic spectral density
$\rho_S(\zeta;\{\mu_i\})$ does have a simple and well-defined power-series
expansion. The explanation is, however, simple. Because of the 
de Wit-`t Hooft poles, we can derive the small-mass expansion of 
$\Sigma_{\nu}(\mu_v,\{\mu_i\})$ up to (and including) order 
$\mu_v^{2N_{f}+2\nu-1}$. But the perturbatively expanded  
$\rho_S(\zeta;\{\mu_i\})$ (see, $e.g.$ the massless case (\ref{rhomassless}))
{\em starts} only at order $\zeta^{2N_{f}+2\nu+1}$. The microscopic spectral
density is always precisely one step ahead of the order to which we can
push the small-mass expansion of the partially quenched chiral condensate!
In hindsight, a phenomenon like this had to occur, since the pure power series
implied by the small-mass expansion cannot give rise to any discontinuity
across the imaginary axis. So the de Wit-`t Hooft poles in fact precisely
save what would otherwise be a paradoxical situation.

\noi
The asymptotic large-mass expansion of section 3.2 cannot be used to extract
the microscopic spectral density either. Here the reason is very different.
Recall that the asymptotic expansion we have derived in section 3.2
is suitable $\mu_v$ on the positive real axis. For example, for $N_f=1$
our expansion simply coincides, up to an irrelevant normalization factor, 
with the asymptotic expansion of the Bessel function $I_0(\mu)$:
\beq
{\cal Z}_0^{(N_f=1)}=I_0(\mu) ~\sim~
\frac{e^{\mu}}{\sqrt{2\pi\mu}}\sum_{k=0}^{\infty}
\frac{(-1)^k}{(2\mu)^k}\frac{\Gamma(k+1/2)}{k!\Gamma(-k+1/2)} 
\label{Besselasymp}
\eeq
This asymptotic expansion is correct, but it neglects exponentially
suppressed terms of order $e^{-\mu}/\sqrt{\mu}$ and lower. Because of
this, the asymptotic expansion (\ref{Besselasymp}) does not 
reproduce the asymptotic expansion for $J_0(\mu)$ after rotation to the
imaginary axis (the exponentially suppressed terms then become of
magnitude comparable to those kept in (\ref{Besselasymp}), and that
expansion is therefore no longer correct near the imaginary axis). 
We conclude that
neither of the two expansions we have considered here are suitable for 
deriving the microscopic spectral density.

%As the partition function of chiral Random Matrix Theory has been shown
%\cite{SV} to
%reduce to (\ref{ZUE}) in the limit where the size of the random matrix
%is large, it must also satisfy (\ref{partdiffeq}) in said limit. In fact the
%uniqueness of the solutions to (\ref{partdiffeq}) offers an alternative way
%to prove the equivalence between the chiral Random Matrix approach and the
%one of the effective theory. For instance, for $N_f=1$  (\ref{partdiffeq})
%gives
%\beq
%\frac{1}{4m}\left\langle \Tr\frac{1}{iD-m} \right\rangle
%-\frac{1}{4}\left\langle \left(\Tr
%\frac{1}{iD-m}\right)^2-\Tr\frac{1}{(iD-m)^2} \right\rangle = 1 ~ .
%\eeq
%Where $D$ is the chRMT analogue of the QCD Dirac operator. The above
%statement is equivalent to
%\beq
%\label{identity?}
%\eeq 
%The function $\tau_2(x,y,m)$ is the connected part of the microscopic
%spectral two-point function,
%$\rho_2(x,y,m)=\rho_S(x,m)\rho_S(y,m)-\tau_2(x,y,m)$. Inserting the universal
%chRMT results found in [Damgaard and Nishigaki NPB 518 (1998) 495] we see that
%the
%identity (\ref{identity?}) is satisfied/violated ?  For $m$ large one can
%also make the somewhat simpler check; using the decoupling of the heavy quark 
%one has $\rho_S(u)=\frac{u}{2}\left[J_{\nu}(u)^2 - J_{\nu+1}(u)J_{\nu-1}(u)
%\right]$ and $\rho_2(x,y)=xy(\frac{xJ_1(x)J_0(y)-yJ_0(x)J_1(y)}{x^2-y^2})^2$
%the identity (\ref{identity?}) is (of course?) still satisfied/violated ?  

\setcounter{equation}{0}
\section{Conclusions}

Using the Virasoro constraints on the effective partition function we have
obtained small-mass and large-mass expansions for the QCD partition function 
valid in the scaling region $V\ll 1/m_\pi^4$. In these expansions we 
can treat the number of valence quarks $N_v$ as a continuous parameter. 
This form is thus suited for a calculation of the partially quenched chiral 
condensate using the replica method.  Two series of spectral sum rules 
for the QCD Dirac operator follow
from the small-mass expansion. One series extends the Leutwyler-Smilga 
sum rules in a simple way to very high (here 
$8^{\rm th}$) order, while the other series is new: it computes the
spectral sum rules in the massive theory, and we have given series
expansions in the physical masses for these new sum rules. In all cases
we have checked, we have found complete agreement with earlier results based
either on Random Matrix Theory or the supersymmetric technique. 
The replica method constitutes a new and independent derivation of
these results.

\noi
With the expansions of the partition function at hand we have also
derived the small-mass and large-mass expansions of the partially
quenched chiral condensate. In all cases we have checked there is again
complete agreement with earlier analytical predictions based on the two other
methods. 

\noi
We have here restricted ourselves to small-mass and large-mass series 
solutions only because they provide the simplest framework in which to employ
the replica method. This is not an inherent restriction, though. It is
thus an open challenge to extend the method beyond these two series
expansions.

\noi
{\sc Acknowledgement:}\\
We thank M. L\"{u}scher and H.B. Nielsen for an early conversation.
This work was supported in part by EU TMR grant no. ERBFMRXCT97-0122.

\end{document}